\documentclass[12pt,a4paper]{article}
\usepackage{amsmath}
\usepackage{amssymb}
\usepackage{graphicx}
\usepackage{verbatim}

\begin{document}
\title{\bf $B\to X_s+{\rm Missing~Energy}$ in unparticle model}
\author{\vspace{0.8cm}\\Yunfei Wu\thanks{yunfei\_wu@pku.edu.cn}\ \ and\ \ Da-Xin Zhang\thanks{dxzhang@mail.phy.pku.edu.cn}\\\small{
School of Physics, Peking University,  Beijing 100871, China}}
\date{}
\begin{titlepage}
\maketitle \thispagestyle{empty} \vspace{1.5cm}
\begin{abstract}
We analyze the inclusive decay mode $B\to X_s+{\rm Missing~Energy}$
in the unparticle model, where an unparticle can also serve as the
missing energy. We use the Heavy Quark Effective Theory in the
calculation. The analytical result of the decay width in the free
quark limit and that of the differential decay rate to the order of
$1/m_b^2$ are presented. Numerical results of the inclusive mode
show interesting differences from those of the exclusive modes. Near
the lower endpoint region, the $d_{\mathcal U}<2$ unparticle has
very different behavior from the Standard Model particles.
\end{abstract}
\end{titlepage}
\section{Introduction}
The rare decay $B\to X_s+\nu\bar\nu$ is very small in the Standard
Model (SM)\cite{QCD} so that it might be very sensitive to the new
physics beyond the SM. Furthermore, in the final state the missing
energy carried by the neutrino-anti-neutrino pair might be polluted
experimentally  by other missing-energy-like states in the presence
of new physics.

Recently, the Unparticle Model suggested by Georgi\cite{unpar}
provides a possible candidate for the missing energy in $B\to
X_s+{\rm Missing~Energy}$. In the Unparticle Model, below a scale
$\Lambda_\mathcal{U}$ the interaction of the unparticle with the  SM
sector takes a form like\cite{unpar}
\begin{equation}
\frac{\mathcal{C}_\mathcal{U}\Lambda_\mathcal{U}^{d_\mathcal{BZ}-d_\mathcal{U}}}{M_\mathcal{U}^k}\mathcal{O}_{SM}\mathcal{O}_\mathcal{U},
\end{equation}
where $\mathcal {C}_\mathcal{U}$ is a coefficient function and $M_{\mathcal U}$ is a large mass scale of the particles mediating the interaction between the SM fields and the unparticle fields. If
$M_\mathcal{U}$ is large enough, the unparticle stuff does not couple
strongly to the ordinary particles. People have introduced many
kinds of couplings\cite{unpar,inter,inter2}, including the Yukawa
and the partial differential couplings. A lot of works have been
done to study the possible consequence of the unparticle, most of
which focus on the exclusive processes. Here we will discuss the
inclusive mode $B\to X_s+{\rm Missing~Energy}$, treating the
unparticle as part of the missing energy. We will use Heavy Quark
Effective Theory (HQET) in our analysis. The results are constrained
by the data \cite{pdg}.


The organization of this paper is as follows. At first, we apply the
HQET results to unparticle model and  give out the general form of
$B\to X_s+\mathcal{U}$ decay rates. Then we present  in the
analytical forms  the decay width in the free quark approximation
and the differential decay width versus the missing energy to the
$1/m_b^2$ order. Numerical results will be given. We will summarize
at the end.

\section{HQET application in Unparticle Model}
In the  Unparticle Model,
the effective Hamiltonian for $B\to
X_s +\mathcal{U}$ at quark level is given by
\begin{equation}
\mathcal {H}_{eff}^\mathcal{S}=\frac{\mathcal {C}_\mathcal
{S}^q\Lambda_\mathcal {U}^{k-d_\mathcal {U}}}{M_\mathcal {U}^k}(\bar
b{\gamma_\mu(1-\gamma_5)}s)\partial^{\,\mu}\mathcal {O}_\mathcal {U}
\end{equation}
for the scalar unparticle, or
\begin{equation}
\mathcal {H}_{eff}^\mathcal{V}=\frac{\mathcal {C}_\mathcal
{V}^q\Lambda_\mathcal {U}^{k+1-d_\mathcal {U}}}{M_\mathcal
{U}^k}(\bar b{\gamma_\mu(1-\gamma_5)}s)\mathcal {O}_\mathcal{U}^{\,\mu}
\end{equation}
for the vector unparticle. Here $\mathcal {C}_\mathcal
{S}^q$ and $\mathcal {C}_\mathcal
{V}^q$ are the dimensionless coupling constant between the quark current and unparticle fields.
The $d_{\mathcal U}$ is a non-integral number severing as the
dimension of unparticle operators. The $d_{\mathcal U}$ can not be small than 1 for the
unitary of the theory\cite{unpar}. If one impose the conformal symmetry on the vector unparticle fields, the primary,
gauge invariant vector unparticle operators could only have dimension $d_{\mathcal U}>3$\cite{comment,comment2}.

We will also introduce the two dimensional
coefficients corresponding to scalar and vector unparticles
\begin{equation}
\textit{c}_\mathcal{S}^{\,q}=\frac{\mathcal {C}_{\mathcal
S}^{q}\Lambda_\mathcal {U}^{k-d_\mathcal {U}}}{M_\mathcal {U}^k}\,,
\quad \text{and} \quad \,\textit{c}_\mathcal{V}^{\,q}=\frac{\mathcal
{C}_{\mathcal V}^{q}\Lambda_\mathcal {U}^{k+1-d_\mathcal
{U}}}{M_\mathcal {U}^k}\,.\label{cq}
\end{equation}

The most general forms of differential decay rates for the scalar
and vector unparticle final states are
\begin{eqnarray} d\Gamma^{\mathcal{S(V)}}&=&\sum_{X_s}(2\pi)^4
\delta^4(p_B-p_{\mathcal{U}}-p_{X_s})\frac{\big|\langle
X_s~\mathcal{U}|H_{eff}^{\mathcal{S(V)}}|B
\rangle\big|^2}{2m_B}\nonumber\\
&& \times
A_{d_\mathcal{U}}\theta(p_\mathcal{U}^0)\theta(p_\mathcal{U}^2)
(p_\mathcal{U}^2)^{d_\mathcal{U}-2}\frac{d^4p_\mathcal{U}}{(2\pi)^4},\label{md}
\end{eqnarray}
where  $A_{d_{\mathcal{U}}}$ is defined as\cite{unpar}
\begin{equation}
A_{d_{\mathcal{U}}}=\frac{16\pi^{5/2}}{(2\pi)^{2d_\mathcal{U}}}\frac{\Gamma(d_\mathcal{U}+1/2)}{\Gamma(d_\mathcal{U}-1)\Gamma(2d_\mathcal{U})}.
\end{equation}
The factor
$A_{d_\mathcal{U}}\theta(p_\mathcal{U}^0)\theta(p_\mathcal{U}^2)
(p_\mathcal{U}^2)^{d_\mathcal{U}-2}$ in (\ref{md}) counts for the
unparticle phase space\cite{unpar}, over which the integrations can
be performed in the rest frame of the $B$ meson. The rest part in
(\ref{md}) containing the matrix element squared are conventionally
written, in analogy with those in the semileptonic decays in the SM,
as the product of the hadron and the unparticle tensors analytical
\cite{frame,hqet,shape,shape2,Sh},
\begin{eqnarray}
&&\sum_{X_s}(2\pi)^3
\delta^4(p_B-p_{\mathcal{U}}-p_{X_s})\frac{\big|\langle
X_s~\mathcal{U}|H_{eff}^{\mathcal{S(V)}}|B
\rangle\big|^2}{2m_B}\nonumber\\
&=&\textit{c}_\mathcal{S(V)}^{\,q\,2}W_{\alpha\beta}U_{\mathcal{S(V)}}^{\alpha\beta},\label{smatrix}
\end{eqnarray}
where the unparticle tensors are
\begin{equation}
U_{\mathcal S}^{\alpha\beta}=p_{\mathcal U}^\alpha p_{\mathcal
U}^\beta\ \ \text{and} \ \ U_{\mathcal
V}^{\alpha\beta}=-g^{\alpha\beta}+\frac{p_{\mathcal U}^\alpha
p_{\mathcal U}^\beta}{p_{\mathcal U}^2},\label{umatrix}
\end{equation}
and the hadronic tensor is defined by
\begin{equation}
W_{\alpha\beta}=\sum_{X_s}(2\pi)^3\delta^4(p_B-q-p_{X_s})\frac{\langle
B(p_B)|J^\dag_\alpha|X_s(p_{X_s})\rangle\langle
X_s(p_{X_s})|J_\beta|B(p_B)\rangle }{2m_B},\label{Wd}
\end{equation}
with $J_\alpha=\bar s \gamma_\alpha(\frac{1-\gamma_5}{2}){b}$.

The most general form of $W_{\alpha\beta}$ is\cite{hqet}
\begin{equation}
W_{\alpha\beta}=-g_{\alpha\beta}W_1+v_\alpha v_\beta
W_2-i\epsilon_{\alpha\beta\gamma\delta}v^\gamma q^\delta
W_3+q_\alpha q_\beta W_4+(v_\alpha q_\beta+q_\alpha
v_\beta)W_5.\label{W}
\end{equation}
The scalar structure functions $W_j$ are functions of $q^2$ and
$v\cdot q$, where $v$ is the four velocity of the heavy bottom
quark.
They  are related to $T_j$'s by applying $W_j=-1/\pi \text{Im} T_j$,
where
\begin{equation} T_{\alpha\beta}=-i\int d^4xe^{-iq\cdot
x}\frac{\langle B|T[J^\dag_\alpha(x)J_\beta(0)]|B\rangle }{2m_B}.
\end{equation}
In general,
\begin{equation}
T_{\alpha\beta}=-g_{\alpha\beta}T_1+v_\alpha v_\beta
T_2-i\epsilon_{\alpha\beta\gamma\delta}v^\gamma q^\delta
T_3+q_\alpha q_\beta T_4+(v_\alpha q_\beta+q_\alpha v_\beta)T_5.
\end{equation}
The HQET provides a systematical tool in investigating the
Heavy-light hadrons such like the $B$ mesons\cite{frame,hqet}.
$T_j$'s can be expanded in $1/m_b$ using HQET, their forms  to
$1/m_b^2$ can be found in Ref.\cite{shape2,Sh}. There emerge some
problems near the endpoint region of the energy
spectrum\cite{shape}, which can be avoided by introducing suitable
cuts in this work.

 Applying Eqs.(\ref{smatrix},\ref{umatrix},\ref{W}) we get
\begin{eqnarray}
d\Gamma^{\mathcal{S}}&=&8\pi{\textit{c}_\mathcal{S}^{\,q}}^2\left[-{p_\mathcal{U}^2}W_1+(v\cdot
p_\mathcal{U})^2W_2+(p_\mathcal{U}^2)^2W_4+2(v\cdot
p_\mathcal{U}){p_\mathcal{U}^2}W_5\right]\nonumber\\
&&\times
A_{d_\mathcal{U}}\theta(p_\mathcal{U}^0)\theta(p_\mathcal{U}^2)
(p_\mathcal{U}^2)^{d_\mathcal{U}-2}\frac{d^4p_\mathcal{U}}{(2\pi)^4},\label{scalar}
\end{eqnarray}
and
\begin{eqnarray}
d\Gamma^{\mathcal{V}}=8\pi{\textit{c}_\mathcal{V}^{\,q}}^2\Big[-3W_1+(1-\frac{(v\cdot
p_\mathcal{U})^2}{p_\mathcal{U}^2})W_2\Big]
A_{d_\mathcal{U}}\theta(p_\mathcal{U}^0)\theta(p_\mathcal{U}^2)
(p_\mathcal{U}^2)^{d_\mathcal{U}-2}\frac{d^4p_\mathcal{U}}{(2\pi)^4}.\label{vector}
\end{eqnarray}
The integrations over $d^4p_\mathcal{U}$ in (\ref{scalar}) and
(\ref{vector}) are constrained by the  function
$\theta(p_\mathcal{U}^0)\theta(p_\mathcal{U}^2)$ and by the
condition $p_{\mathcal U}^0<m_B-{m_{X_s}}_{min}$.

\section{Differential decay rates and the width}
The inclusive differential decay rates are calculated  using
(\ref{scalar}) and (\ref{vector}) by taking $m_s \to 0$. Terms with
derivatives of $\delta$ function are evaluated using integrating by
parts.
Then the differential decay width for the scalar unparticle emission
is
\begin{eqnarray}
\frac{d\Gamma^{\mathcal S}}{dx}&=&\frac {A_{d_{\mathcal
U}}{\textit{c}_\mathcal{S}^{\,q}}^2{{ m_b}}^{2\,d_{\mathcal U}-1} (
2\,x-1 ) ^{d_{\mathcal U}-4} }{3\pi^2\,(x-1)} \bigg\{ {\theta}
 ( x-1/2 )  \Big[ 3/2\,{{ m_b}}^{2} ( 2\,
x-1 ) ^{2} ( x-1 ) ^{3}\nonumber\\
&&
+1/2\,{\lambda_1}\, \Big( 2
\,{d_{\mathcal U}}^{2} ( 8\,{x}^{2}-5\,x+1 )  ( x-1 ) ^{4}-d_{\mathcal U}
 ( 48\,{x}^{4}+40\,{x}^{3}-97\,{x}^{2}+47\,x-6 )  ( x-
1 ) ^{2}\nonumber\\
&&
+50\,{x}^{4}+3-18\,x-40\,{x}^{5}+49\,{x}^{2}+32\,{x}^{6}
-66\,{x}^{3} \Big) \nonumber\\
&&
-3/4\,{\lambda_2}\, ( 2\,x-1 )
 \Big( 2\,d_{\mathcal U} ( 37\,{x}^{2}-39\,x+10 )  ( x-1 ) ^
{2}-20\,{x}^{4}-19\,x-8\,{x}^{3}+36\,{x}^{2}+3 \Big)  \Big]  \nonumber\\
&&
+1/4\,
{\delta} ( x-1/2 )  (\ 2\,x-1 )  ( x-1
 ) ^{2} \Big[ {\lambda_1}\, \Big( ( 4\, ( 8\,{x}^{2}-5\,x
+1)  ( x-1) ^{2}d_{\mathcal U}\nonumber\\
&&
+8+135\,{x}^{2}-82\,{x}^{3}-61\,x-
32\,{x}^{4} )
 \Big)
+3\,{\lambda_2}\, ( 2\,x-1 )  ( 37\,{x}^{2}
-39\,x+10 )  \Big] \nonumber\\
&&+1/4\,{\delta'} (x-1/2 ) {
\lambda_1}\, ( 8\,{x}^{2}-5\,x+1 )  ( 2\,x-1
 ) ^{2} ( x-1 ) ^{4} \bigg\} ,{\label{difs}}
\end{eqnarray}
and for the vector unparticle emission, it is
\begin{eqnarray}
\frac{d\Gamma^{\mathcal V}}{dx}&=&\frac { A_{d_{\mathcal
U}}{\textit{c}_\mathcal{V}^{\,q}}^2{m_b}^{2\,d_{\mathcal U} -3} (
2\,x-1 )^{d_{\mathcal U}-5}}{12\pi^2 (x-1)} \bigg\{ {\theta} (x-1/2
)
 \Big[ -6\,{{ m_b}}^{2} ( 3-9\,x+8\,{x}^{2} )  ( 2
\,x-1 ) ^{2} ( x-1 ) ^{2}\nonumber\\
&&
+{\lambda_1}\, \Big( 4\,{
d_{\mathcal U}}^{2} ( 3-9\,x+8\,{x}^{2} )  ( x-1 ) ^{4}+2\,d_{\mathcal U}
 ( 12\,{x}^{3}-19\,{x}^{2}+15\,x-6 )  ( x-1 ) ^{
2}\nonumber\\
&&
-12\,x-128\,{x}^{6}-306\,{x}^{4}+6+316\,{x}^{5}-14\,{x}^{2}+130\,{x}
^{3} \Big) \nonumber\\
&&
+{\lambda_2}\, \Big( 6\,d_{\mathcal U} ( 2\,x-1 )
 ( 40\,{x}^{3}-85\,{x}^{2}+69\,x-18 )  ( x-1 ) ^
{2}\nonumber\\
&&
-3\, ( 2\,x-1 )  ( 160\,{x}^{5}-518\,{x}^{4}+722\,{
x}^{3}-544\,{x}^{2}+225\,x-39 ) \Big)  \Big] \nonumber\\
&&
+{\delta}
 ( x-1/2 )  ( x-1 ) ^{2} ( 2\,x-1 )
 \Big[ {\lambda_1}\, \Big( 4\,d_{\mathcal U} ( 3-9\,x+8\,{x}^{2} )
 ( x-1 ) ^{2} \nonumber\\
&&
+x ( 39\,x-15+16\,{x}^{3}-38\,{x}^{2}
 )  \Big)
+{\lambda_2}\, ( 6\,x-3 )  ( 40\,{
x}^{3}-85\,{x}^{2}+69\,x-18 )  \Big] \nonumber\\
&&
+{\delta'} (x-1
/2 ) {\lambda_1}\, ( x-1 ) ^{4} ( 2\,x-1
 ) ^{2} ( 3-9\,x+8\,{x}^{2} )  \bigg\} ,\label{difv}
\end{eqnarray}
where $\lambda_1$ and $\lambda_2$ are parameters in the heavy quark
expansion. Here we introduce the dimensionless variable
$x=p_{\mathcal U}^0/m_b$, which serves as unparticle energy
modulated by the heavy quark mass.

Further integration over $x$ will not generally give analytical
results for an arbitrary $d_{\mathcal U}$ because of the factor
$(2x-1)^{2d_{\mathcal U}-4({\text{or}}\ 5)}/(x-1)$ in Eq.\ref
{difs}(or \ref {difv}).
However, when taking the limit $\lambda_1=\lambda_2=0$ we can
integrate over $x$. We get
\begin{equation}
\Gamma^{\mathcal{S}}=A_{d_{\mathcal
U}}\frac{{\textit{c}_\mathcal{S}^{\,q}}^2}
{8\pi^{2}}\frac{m_b^{2d_{\mathcal U}+1}}{(d_{\mathcal
U}^2-1)d_{\mathcal U}}, {\label{lsts}}
\end{equation}
and
\begin{eqnarray}
\Gamma^{\mathcal{V}}=\frac{{\textit{c}_\mathcal{V}^{\,q}}^2
A_{d_{\mathcal U}}m_b^{2d_{\mathcal U}-1}(2d_{\mathcal
U}^2-5d_{\mathcal U}+5)}{8\pi^2d_{\mathcal U}(d_{\mathcal
U}-1)(d_{\mathcal U}+1) (d_{\mathcal U}-2)} {\ \ \text{for}\ \
d_{\mathcal U}>2} . \label{lstv}
\end{eqnarray}
Eqs.(\ref {lsts}) and (\ref {lstv}) are the decay widths for  $b\to
s + \mathcal{U}$  in the free quark approximation.

We find that, even if $d_\mathcal{U} = 1$, the scalar unparticle in
the final state gives a finite contribution in (\ref{lsts}), as a
result of the fact that the singularity in the factor  $(d_{\mathcal
U}-1)^{-1}$ is compensated by the factor $A_{d_{\mathcal U}}$. This
is not a common feature as in the exclusive processes such like in
Ref.\cite{us}. We cannot give a simple analytical formula like
(\ref{lstv}) for the vector unparticle in the final state when
$d_{\mathcal U}<2$. But under the assumption of exact conform
symmetry $d_\mathcal{U}$ must be bigger than 3\cite{comment}, then
Eq.(\ref{lstv}) is enough at the first order.

\section{Numerical results}
Numerical results are needed in order to exhibit the unparticle
effects. Both the unparticle and neutrino-anti-neutrino pairs serve
as the missing energy ${\not}E$. We can not distinguish them in the
experiments, so the process $B\to X_s+{\not}E$ may contain both $\nu
\bar{\nu}$ and $ \mathcal {U}$ in the final states. The decay width
is
\begin{equation}
\Gamma(B\to
X_s+{\not}E)=\Gamma(B\to X_s\nu\bar\nu)+\Gamma(B\to X_s\mathcal{U}),\label{dga}
\end{equation}
where $\Gamma(B\to X_s\nu\bar\nu)$ 
comes from the SM and $\Gamma(B\to X_s\mathcal{U})$  come from
either the scalar or the vector unparticle contribution.

Present calculation in SM\cite{QCD,weak} gives
\begin{equation}
B_{SM}(B\to X_s +\nu \bar{\nu})=(3.4\pm0.7)\times10^{-5}.
\end{equation}
The experimental bound\cite{pdg},
\begin{equation}
B_{exp}(B\to X_s+{\not}E)<6.4\times10^{-4},
\end{equation}
is about one order larger than the SM calculation. This large
difference allows new physics to provide candidates as the missing
energy. In the unparticle model the candidate is the unparticle.

There are two kinds of singularities brought by  the unparticle: one
comes from the $(2x-1)^{d_{\mathcal U}-4(5)}$ and the other from
$1/(x-1)$. At the quark level, the endpoint of unparticle energy
spectrum is singular. But the true endpoint are at
$(m_B^2-m_K^2)/2m_B$ and $m_B-m_K$, depending on the masses of the
hadrons. Near $m_B-m_K$ or $x\sim 1$ when the HQET fails, we
introduce some cuts. For the scalar unparticle model, we set the
$x=0.975$, {\it i. e.} we take the $\Lambda_{QCD}\sim 200$MeV that
around $m_B-\Lambda_{QCD}$ the HQET fails\cite{shape}. But for the
vector unparticle theory, the case is different because the
$\theta(1-x)$ term in (\ref{difv}) goes through zero when $x\sim
0.923$, so it severs as our cut. We have taken the mass parameters
as\cite{pdg,ham}
\begin{equation}
m_b=4.7\text{GeV}, m_B=5.279\text{GeV}, \text{and}\
m_K=0.493\text{GeV},
\end{equation}
and the heavy quark expansion parameters as \cite{shapeparame}
\begin{equation}
\lambda_1=-0.497\text
{GeV}^2\ \text{and}\ \lambda_2=0.12\text {GeV}^2\, .
\end{equation}

There are very distinctive differences between the unparticle model
and the SM in the differential decay widths when $d_{\mathcal U}<2$.
Near the lower endpoint region $x\sim (m_B^2-m_K^2)/2m_B m_b$, the
SM differential width approach zero, while the unparticle model
gives finite results. This comes from $(p_{\mathcal
U}^2)^{d_{\mathcal U}-2}$ in the phase space of the unparticle, see
Eqs.(\ref{scalar}) and (\ref{vector}). When $p_{\mathcal U}^2$ goes
to zero, the phase space goes to infinity. But the SM final state
$\nu\bar\nu$ has no such an enhancement. For the vector unparticle
model , the differential width also gives a finite result at $x\sim
(m_B^2-m_K^2)/2m_B m_b$ when $d_{\mathcal U}>2$. Here, the
enhancement comes from the vector unparticle tensor $U_{\mathcal
V}^{\alpha\beta}$ in Eq.(\ref{umatrix}). $p_{\mathcal U}^2$ appears
in the denominator.

Near the endpoint $x\sim (m_B-m_K)/m_b$, the scalar unparticle model
results go to infinity, which comes from the HQET. There are also
such kinds of singularity in SM\cite{hqet,shapeend,gaend}. For the
vector unparticle model, the results go to negative infinity when
$x$ goes to 1. And the vector unparticle contributions vanish near
$x=1$. This turning comes the from the form of matrix element in
(\ref{vector}) and HQET.

We plot the spectra for the scalar unparticle model in Fig. 1 for
$d_{\mathcal U}\leq 2$ and Fig. 2 for $d_{\mathcal U}> 2$,
respectively. The spectra for vector unparticle model  are given in
Fig. 3. for  $d_{\mathcal U}<3$ , and in Fig. 4 for  $d_{\mathcal
U}\geq 3$, respectively. Note that   $d_{\mathcal U}<3$ is allowed
if we give up the strict conformal symmetry.

The branching ratios of scalar and vector unparticle emission
processes show many resemblances if one neglect the conformal
constraint on $d_{\mathcal U}$. They are presented in Fig.
\ref{bscalare} for the scalar unparticle, and in Fig. \ref{vectsma}
and \ref{vectsmd} for the vector unparticle.
Figs. \ref{vectsma} is allowed if one gives up the  strict conformal
symmetry, with the unparticle contribution  peaking at $d_{\mathcal
U}=1.3\sim1.4$. In this case the branching ratio goes down when
$d_{\mathcal U}$ becomes bigger, as is exhibited in (\ref{lsts}) or
(\ref{lstv}). This is quite different from the previous results
gained from the similar exclusive decay processes\cite{us, Bexclu},
where the branching ratios go up as $d_{\mathcal U}$ is increaing.
It is also important to note the different definitions of the
couplings from Ref.\cite{Bexclu}.

\begin{figure}[h]
\raisebox{180pt}{\rotatebox{0}{\large $\frac{d\Gamma^{\mathcal S}}{dx}$}}
\includegraphics[width=300pt]{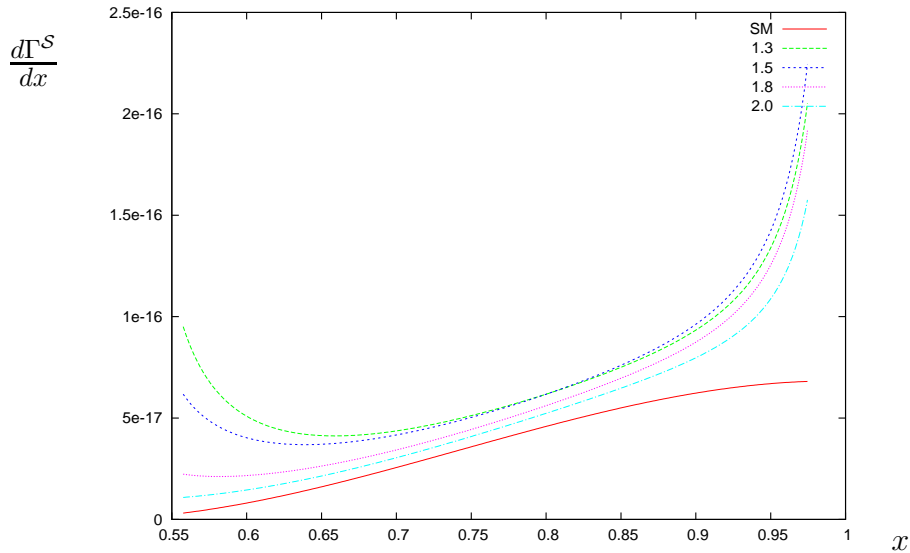}
\makebox(10,0)[br]{$x$} \caption{\footnotesize The scalar unparticle
energy spectrum in inclusive $B\to X_s + {\not}E$ with
${\textit{c}_\mathcal{S}^{\,q}}^2\,=\,1\times10^{-17}$. $d_{\mathcal
U}=1.3, 1.5, 1.8\, \text{and}\, 2$, increasing from top to bottom.
The solid line represents the pure SM result. We cut at $x=0.975$.}
\end{figure}

\begin{figure}[h]
\raisebox{180pt}{\rotatebox{0}{\large $\frac{d\Gamma^{\mathcal S}}{dx}$}}
\includegraphics[width=300pt]{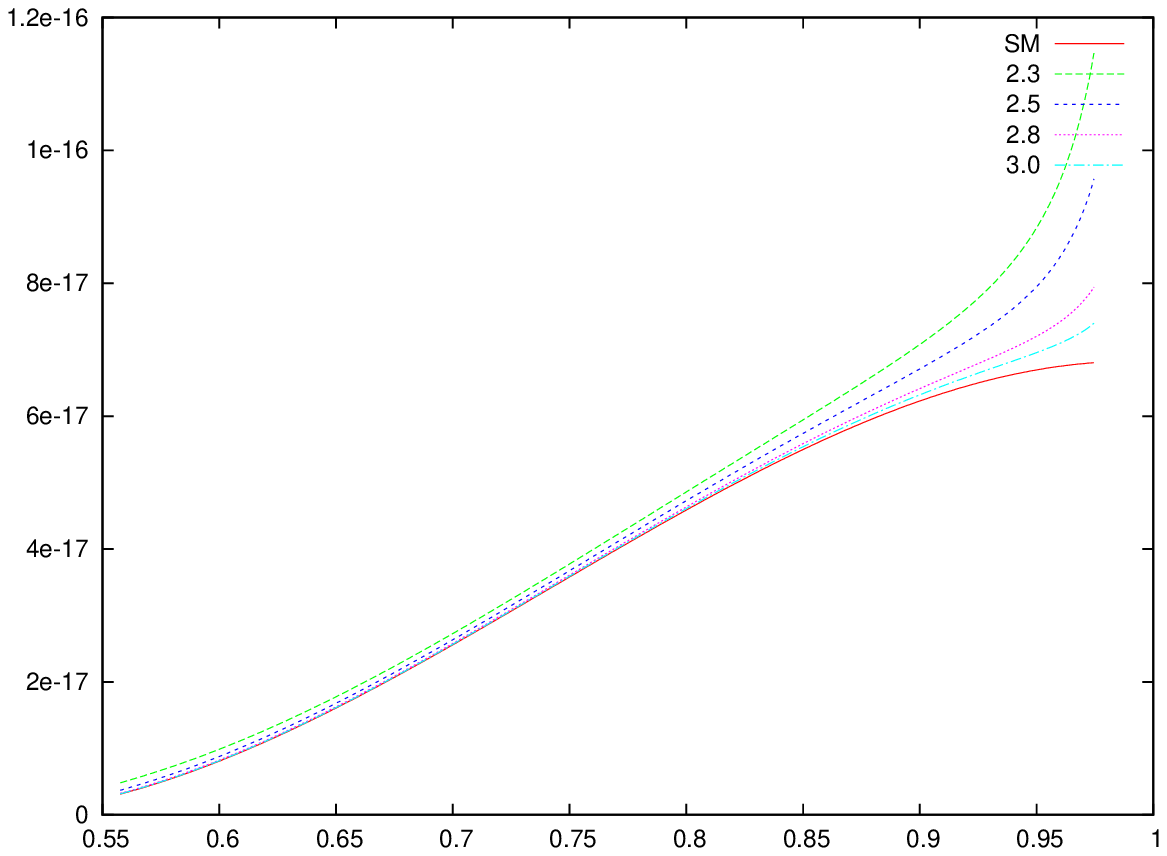}
\makebox(10,0)[br]{$x$} \caption{\footnotesize The scalar unparticle
energy spectrum in inclusive $B\to X_s + {\not}E$ with
${\textit{c}_\mathcal{S}^{\,q}}^2\,=\,1\times10^{-17}$. $d_{\mathcal
U}=2.3, 2.5, 2.8\, \text{and}\, 3$, increasing from top to bottom.
The solid line represents the pure SM result. The cut is at
$x=0.975$.}
\end{figure}

\begin{figure}[h]
\raisebox{180pt}{\rotatebox{0}{\large $\frac{d\Gamma^{\mathcal V}}{dx}$}}
\includegraphics[width=300pt]{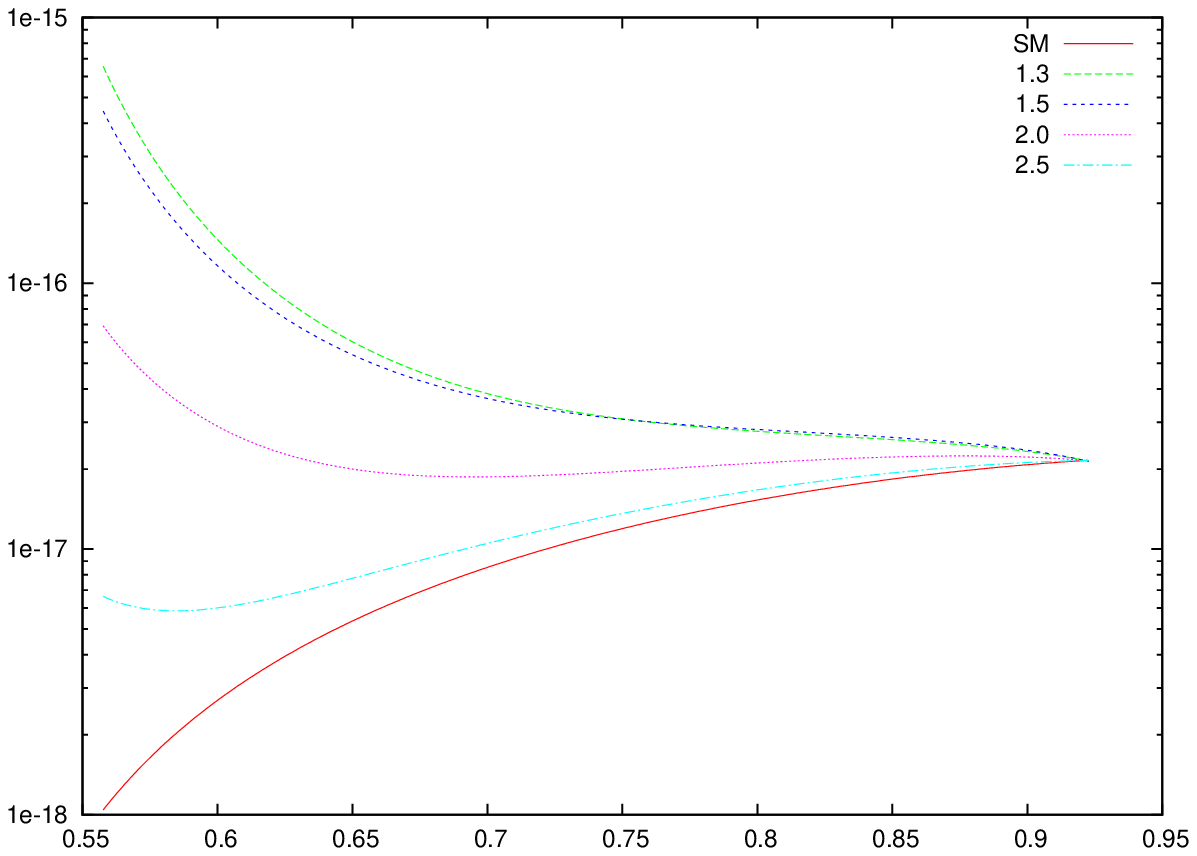}
\makebox(10,0)[br]{$x$} \caption{\footnotesize The vector unparticle
energy spectrum in inclusive $B\to X_s + {\not}E$ with
${\textit{c}_\mathcal{V}^{\,q}}^2\,=\,1\times10^{-15}$ .
$d_{\mathcal U}=1.3\, 1.5\, 2\, \text{and}\, 2.5$, increasing from
top to bottom. The solid line represents the pure SM result. The cut
is at $x=0.923$.} {\label{vectd}}
\end{figure}

\begin{figure}[h]
\raisebox{180pt}{\rotatebox{0}{\large $\frac{d\Gamma^{\mathcal V}}{dx}$}}
\includegraphics[width=300pt]{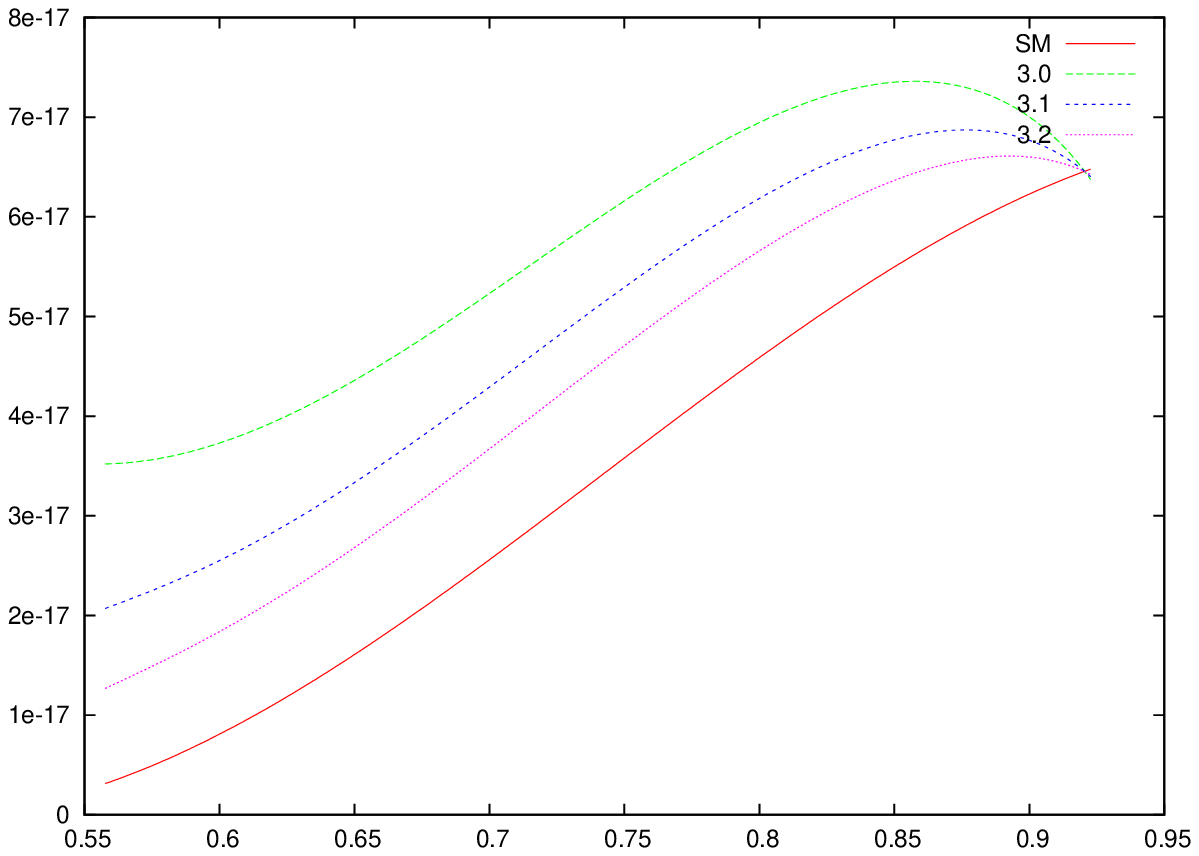}
\makebox(10,0)[br]{$x$} \caption{\footnotesize The vector unparticle
energy spectrum in inclusive $B\to X_s + {\not}E$ with
${\textit{c}_\mathcal{V}^{\,q}}^2\,=\,1\times10^{-14}$. $d_{\mathcal
U}=3.0, 3.1, \, \text{and}\, 3.2$, increasing from top to bottom.
The solid line represents the pure SM result. The cut is at
$x=0.923$.}
\end{figure}

\begin{figure}[h]
\raisebox{100pt}{\rotatebox{90}{{\large$\frac{\pi^2}{{\textit{c}_\mathcal{S}^{\,q}}^2}$}{\footnotesize$\Gamma(B\to
X_s + \mathcal{U})$}}}
\includegraphics[width=300pt]{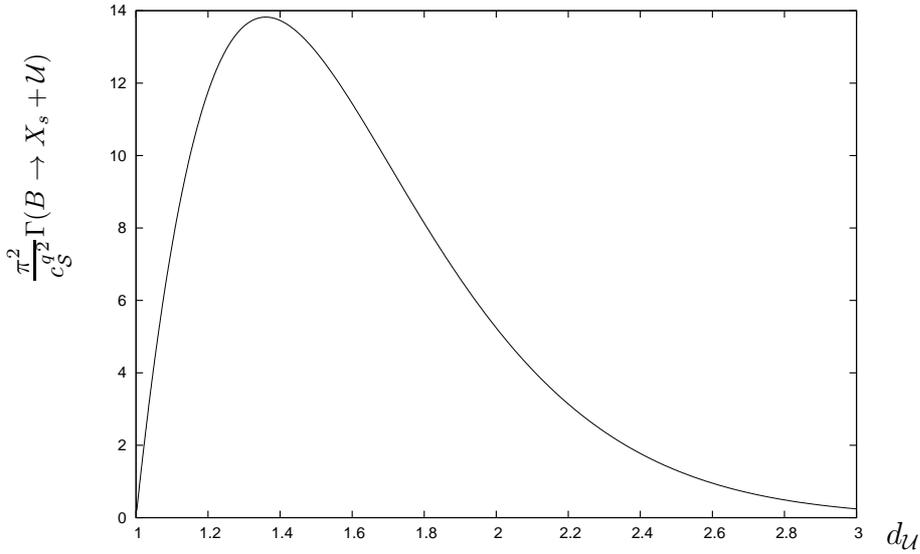}
\makebox(10,0)[br]{$d_{\mathcal U}$} \caption{\footnotesize The
decay width versus $d_\mathcal{U}$ in the scalar unparticle model.
It is  modulated by
${{\textit{c}_\mathcal{S}^{\,q}}^2}/{\pi^2}$.}\label{bscalare}
\end{figure}

\begin{figure}[h]
\raisebox{100pt}{\rotatebox{90}{{\large$\frac{\pi^2}{{\textit{c}_\mathcal{V}^{\,q}}^2}$}{\footnotesize$\Gamma(B\to X_s + \mathcal{U})$}}}
\includegraphics[width=300pt]{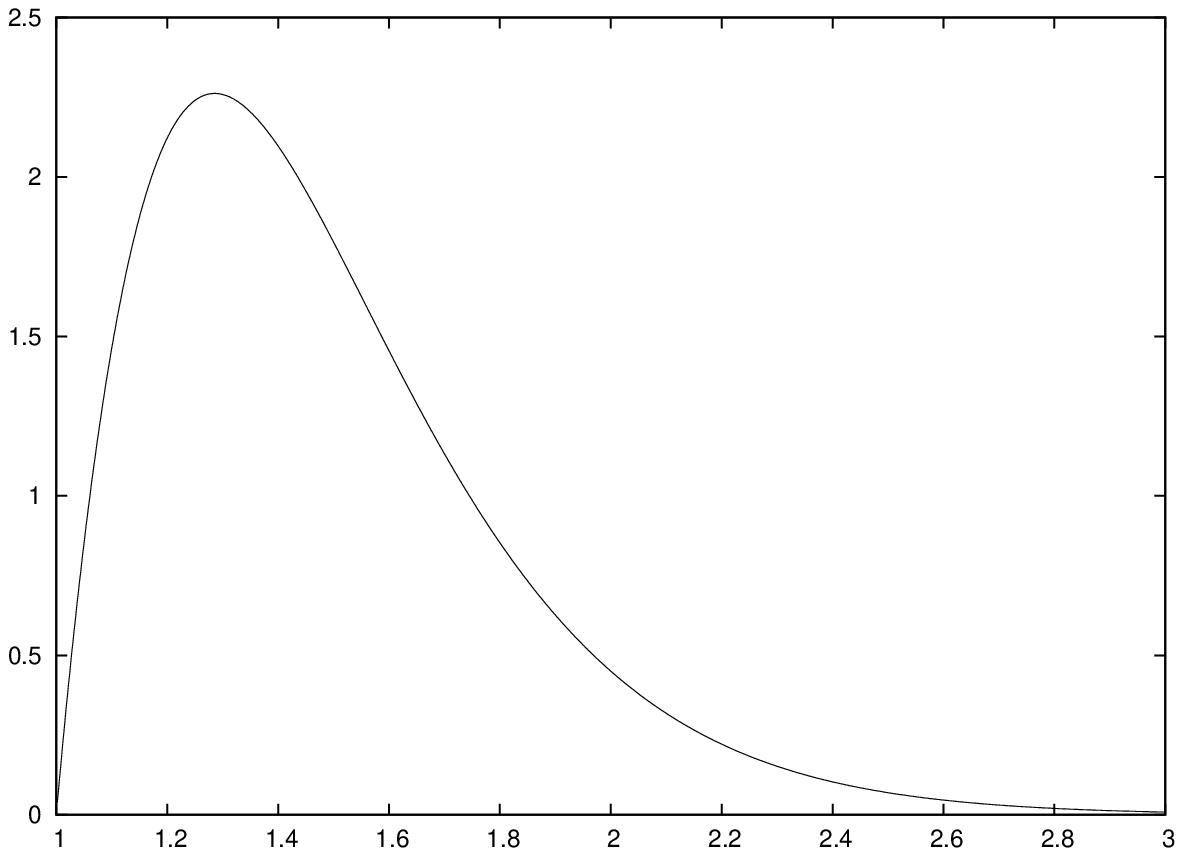}
\makebox(10,0)[br]{$d_{\mathcal U}$} \caption{\footnotesize The
decay width versus $d_\mathcal{U}$ in the vector unparticle model
for  $d_{\mathcal U}<3$. }\label{vectsma}
\end{figure}

\begin{figure}[h]
\raisebox{100pt}{\rotatebox{90}{{\large$\frac{\pi^2}{{\textit{c}_\mathcal{V}^{\,q}}^2}$}{\footnotesize$\Gamma(B\to X_s + \mathcal{U})$}}}
\includegraphics[width=300pt]{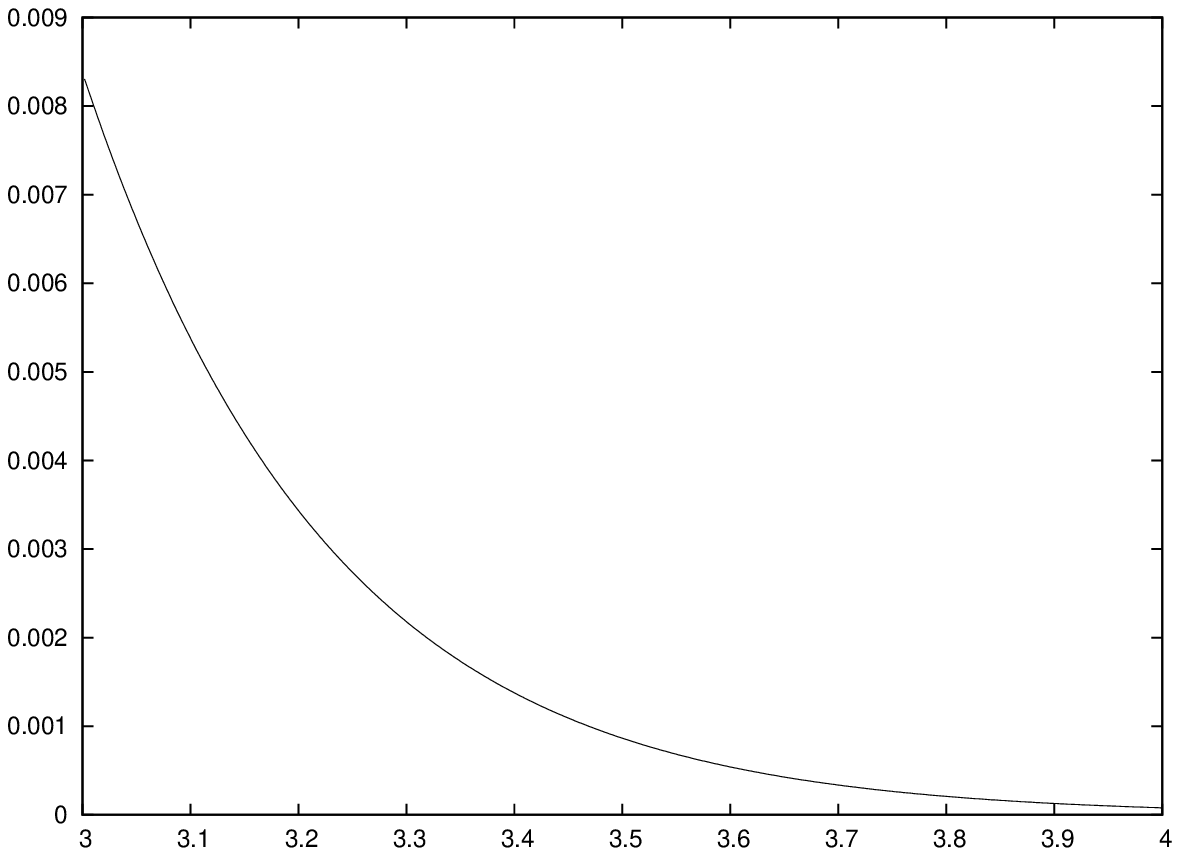}
\makebox(10,0)[br]{$d_{\mathcal U}$} \caption{\footnotesize The
decay width versus $d_\mathcal{U}$ in the vector unparticle model
for  $d_{\mathcal U}\geq 3$ .}\label{vectsmd}
\end{figure}

\section{Summary}

In this paper we have discussed the process $B\to X_s +$Missing
Energy in the unparticle model and given some analytical results of
the decay withes in free quark limit and the differential decay
rates to the $1/m_b^2$ order.  If $d_{\mathcal U}=1.3\sim1.4$, the
unparticle stuffs is most likely to be tested.
If one regards the conformal symmetry\cite{comment}, the vector
unparticle has the most distinctive effect around $d_{\mathcal
U}=3$. Near the lower endpoint region $x\sim (m_B^2-m_K^2)/2m_B m_b$
in the spectrum, the unparticle model show very distinctive behavior
from the SM. This is very possible to be tested in experiments.

This work was supported in part by the National Natural Science
Foundation of China (NSFC) under the grant No. 10435040.

\end{document}